\documentclass[baaa]{baaa}

\usepackage[pdftex]{hyperref}
\usepackage{subfigure}
\usepackage{natbib}
\usepackage{helvet,soul}
\usepackage[font=small]{caption}

\contriblanguage{0}

\contribtype{2}

\thematicarea{11}

\title{Estudio arqueoastronómico en la costa norcentral del Perú}

\titlerunning{Estudio arqueoastronómico en la costa norcentral del Perú}

\author{
J. Ricra\inst{1,2} \& A. Gangui\inst{3}
}

\authorrunning{Ricra \& Gangui}

\contact{jricram@uni.pe}

\institute{
Grupo Astronom\'ia, Facultad de Ciencias, Universidad Nacional de Ingenier\'ia, Per\'u \and
Observatorio Astron\'omico AFARI, Per\'u \and
Instituto de Astronom\'ia y F\'isica del Espacio, CONICET--UBA, Argentina
}

\resumen{
La civilización Caral se desarrolló en la costa norcentral del Perú y tuvo un periodo de ocupación comprendido entre los años 2870 y 1970 a.C. Los primeros estudios realizados en el campo de la arqueoastronomía mostraron evidencia de posibles orientaciones astronómicas en algunos edificios de su ciudad capital, la Ciudad Sagrada de Caral. Sin embargo, cuestiones metodológicas ponen en duda estas conclusiones. Un reciente estudio realizó un análisis estadístico más general, el cual abarcó un total de 55 estructuras arquitectónicas distribuidas en diez asentamientos urbanos que formaron parte de esta civilización, logrando así identificar patrones de orientación de tipo topográfico y astronómico. Partiendo de esta evidencia, planteamos realizar un nuevo estudio centrado de la ciudad capital, con el objetivo de analizar el patrón de orientación de la ciudad, poniendo énfasis en el análisis de las estructuras de mayor importancia religiosa y administrativa con el fin de determinar su funcionalidad y sus posibles vínculos con objetos astronómicos de relevancia. El estudio incluirá un trabajo de campo para la medición de las diversas estructuras y el posterior análisis estadístico de los datos, empleando histogramas de declinación, funciones de densidad y test de probabilidad.}

\abstract{
The Caral civilization developed on the north-central coast of Peru and had an occupation period between 2870 and 1970 years BC. The first studies carried out in the field of archaeoastronomy showed evidence of possible astronomical orientations in some buildings of its capital city, the Ciudad Sagrada de Caral. However, methodological issues cast doubt on these conclusions. A recent study carried out a more general statistical analysis, which covered a total of 55 architectural structures distributed in ten urban settlements that were part of this civilization, thus managing to identify topographic and astronomical orientation patterns. Based on this evidence, we propose to carry out a new study focused on the capital city, with the objective of analyzing the orientation pattern of the city, placing emphasis on the analysis of the most important religious and administrative structures in order to determine their functionality and their possible links with relevant astronomical objects. The study will include field work to measure the various structures and the subsequent statistical analysis of the data, using declination histograms, density functions and probability tests.
}


\begin{document}

\maketitle

\section{Introducción}\label{S_intro}
Una práctica común en diversas civilizaciones antiguas fue la orientación astronómica de sus estructuras arquitectónicas de mayor relevancia. En ciertos casos, la orientación solo tenía como fin el culto a una deidad del cielo, mientras que en otros, era parte de un sistema calendárico que regía las actividades productivas y las costumbres religiosas de la sociedad. Entre los casos más representativos se hallan Göbekli Tepe \citep{Collins2014}, el Círculo de Goseck \citep{Brown2016}, Playa Nabta \citep{Malville} y Stonehenge \citep{HAWKINS1963}.

La orientación de templos o estructuras fue una práctica que también se llevó a cabo en el continente americano. Uno de los casos más tempranos (aunque aún en debate) es el de la cultura Valdivia, en el Ecuador costero (4400-1600 a.C.). En el sitio arqueológico de Real Alto se ha identificado una estructura ceremonial (estructura 7) orientada hacia la salida del Sol en el solsticio de junio \citep{Zeidler1998}. En el territorio peruano encontramos el sitio arqueológico de Buena Vista (2200-2100 a.C.) en el valle del río Chillón, contemporáneo con el periodo final de Caral \citep{Shady2017}. \cite{AdkinsBenfer2009} así como \cite{Benfer2007} sugieren que el Templo del Zorro se encuentra orientado hacia la salida del Sol en el solsticio de diciembre. Además, proponen la existencia de marcadores de horizonte que habrían servido para determinar las posiciones extremas de la Luna (lunasticios). 

La civilización Caral se desarrolló en la zona norcentral del Perú y tuvo un período de ocupación comprendido entre los años 2870 y 1970 a.C. \citep{Solis2001}. A diferencia de los casos anteriores, la civilización Caral tuvo un área de influencia considerable, en donde se han identificado 25 asentamientos urbanos con arquitectura monumental, siendo la Ciudad Sagrada de Caral su ciudad capital \citep{Shady2006}.

Entre los primeros estudios hechos en el campo de la arqueoastronomía en Caral resaltan los de \cite{Pinasco2004} y \cite{Marroquin2010}. Ambos estudios utilizaron planos arqueológicos para identificar orientaciones astronómicas en los edificios que conforman la Ciudad Sagrada de Caral. \cite{Pinasco2004} logró identificar orientaciones solsticiales, lunares y estelares, mientras que \cite{Marroquin2010} identificó orientaciones solsticiales en algunas de las principales pirámides de la ciudad. A pesar de que ambos estudios arrojaron resultados interesantes, estos son cuestionados, por ejemplo, porque la metodología empleada no tuvo en consideración la altura irregular del horizonte que rodea al sitio, imprescindible para el correcto cálculo de las declinaciones astronómicas de aquellos objetos del cielo que podrían ser objeto de las orientaciones.

Un estudio reciente \citep{Gonzalez-Garcia2021}, ofrece una estadística global donde se identifican patrones de orientación topográfico (hacia el río Supe) y astronómico (hacia la salida de la Luna durante el lunasticio mayor sur y hacia la salida del Sol durante el solsticio de diciembre) en una muestra de 55 estructuras arquitectónicas pertenecientes a diez asentamientos (nueve en el valle del río Supe y uno en el valle del río Huaura). Por su naturaleza estadística el estudio no se enfoca en casos particulares, quedando pendiente el estudio del tipo de orientación que poseen los edificios de mayor importancia religiosa o administrativa. 

\section{Hipótesis}

Un problema recurrente en sitios arqueológicos como Göbekli Tepe, el Círculo de Goseck, Playa Nabta, Stonehenge y Valdivia, es su singularidad. Es decir, no hay otro monumento en el entorno de cada sitio arqueológico que reproduzca las mismas orientaciones astronómicas. Por lo tanto, en acuerdo con la premisa \textit{testis unus, testis nullus}, no estamos en condiciones de hacer afirmaciones categóricas sobre la intencionalidad de sus constructores. Pues, un único caso de algo (un experimento, una prueba, una construcción antigua) no es indicio suficiente para elaborar una teoría, puesto que no es falsable. 
Investigaciones recientes han implementado una nueva metodología basada en análisis estadísticos y pruebas de confiabilidad que, por sus características, solo son aplicables a un número significativo de monumentos pertenecientes a un mismo espacio cultural. Como ejemplo, se puede mencionar el caso de los templos del Alto Egipto y Bajo Nubia \citep{Shaltout2005}, Petra \citep{Belmonte2013} o las iglesias cristianas de diversos sitios de América y Europa \citep{Gangui2016}, como en Canarias \citep{Gangui2016a,Paolo2020}. 

\begin{figure}[!t]
\centering
\includegraphics[width=\columnwidth]{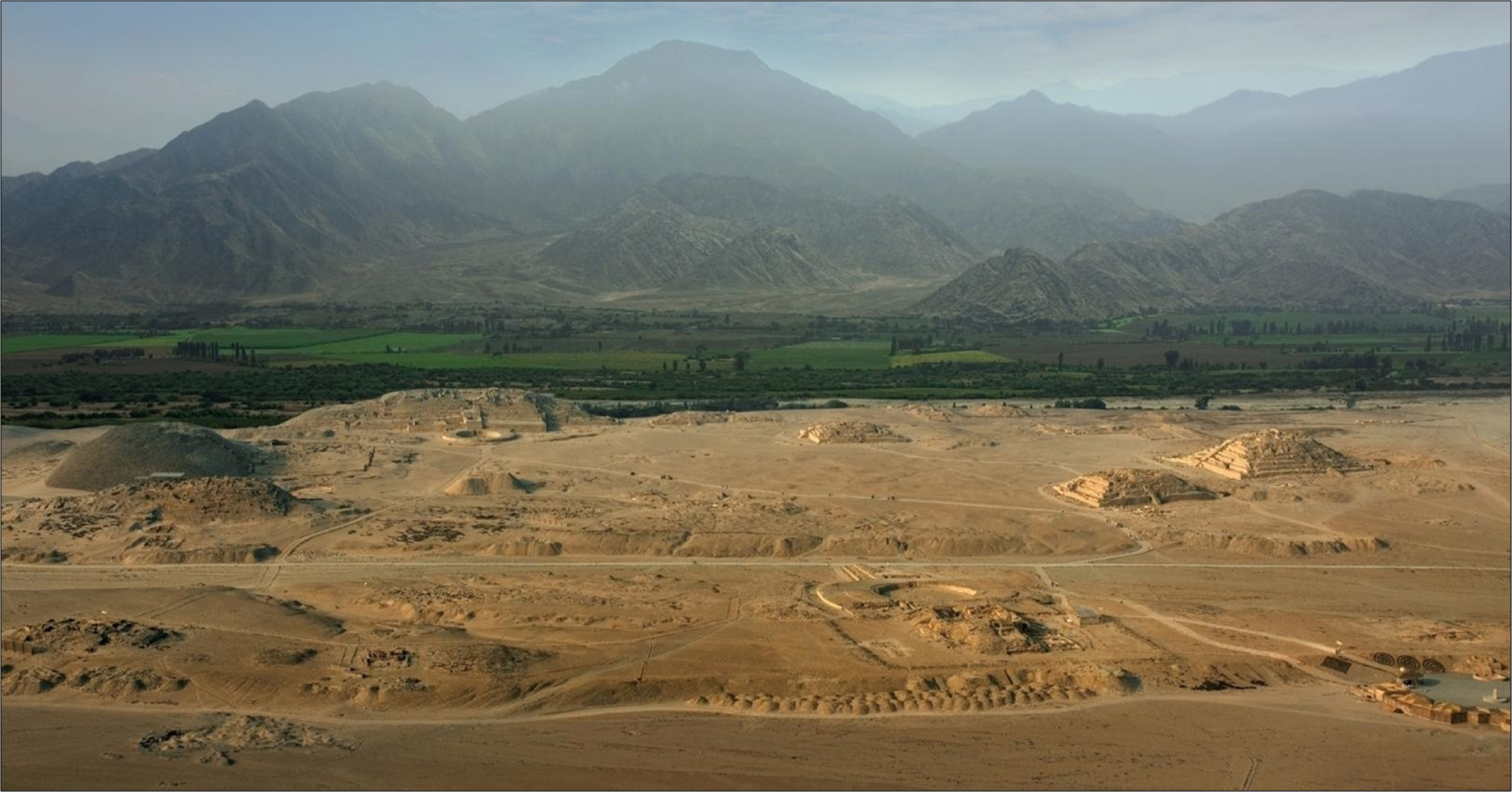}
\caption{Zona nuclear de la Ciudad Sagrada de Caral. Figura extraida de \cite{Shady2017}.}
\label{Figura}
\end{figure}

Por la cantidad de monumentos, Caral reúne las condiciones ideales para la aplicación de un estudio estadístico (como se muestra en \cite{Gonzalez-Garcia2021}). Este mismo tipo de análisis puede ser implementado en su ciudad capital, la cual cuenta con cerca de cincuenta estructuras arquitectónicas y alberga a los templos y pirámides más importantes de esta civilización (ver Fig.~\ref{Figura}). Además, puede ser el preámbulo para un análisis más detallado del tipo de orientación, el nivel de precisión alcanzado y la funcionalidad de los templos y pirámides de mayor importancia religiosa y administrativa. 
Por último, la ya mencionada omisión de las medidas de altura del horizonte y de exploración de posibles marcadores astronómicos, en los trabajos previos \citep{Pinasco2004,Marroquin2010}, hace necesario un nuevo estudio que incorpore en el análisis estos elementos, en particular, porque los cerros que rodean el sitio arqueológico presentan ángulos de elevación considerables. 

\section{Metodología}\label{sec:guia}

\subsection{Cálculo de la posición de objetos celestes para la fecha de estudio}

El periodo de ocupación de la civilización Caral se encuentra entre los años 2870-1970 a.C. \citep{Solis2001,Shady2006}. Debido a su antigüedad, es necesario calcular la posición precisa de los objetos astronómicos más relevantes del cielo, que se verá modificada para cada etapa de su construcción a lo largo de nueve siglos.
Para el caso de los solsticios, la declinación solar es obtenida del valor del ángulo de oblicuidad, que es determinado por la oblicuidad media y la corrección por nutación en la oblicuidad. La oblicuidad media se puede obtener a partir de la fórmula de \cite{Laskar1986}, mientras que para la nutación en la oblicuidad se implementará el modelo IAU2000B desarrollado por \cite{McCarthy2003}. Para el caso de los lunasticios mayores y menores, la declinación lunar se obtendrá a partir del ángulo de oblicuidad y del ángulo de inclinación medio de la órbita lunar respecto a la eclíptica \citep{Williams2003}, teniendo en cuenta, además, el efecto de paralaje lunar \citep{Ruggles1999}.

Para estudiar la posible orientación con estrellas, se realizarán estudios que incluyan la corrección por movimiento propio y precesión. La corrección por movimiento propio se puede realizar a partir de los datos disponibles en la base de datos SIMBAD, mientras que la corrección por precesión se llevará a cabo mediante el empleo del modelo IAU76 desarrollado por \cite{Lieske1977}.

\subsection{Trabajo de campo}

Los datos de las orientaciones de las estructuras bajo estudio deben obtenerse \textit{in situ}, y en términos generales se llevan a cabo de la manera siguiente: se toma el acimut de la dirección que se desea medir y la altura angular del horizonte (\textit{h}) visible desde la ubicación de la construcción. En los casos simples se utiliza un tándem de brújula de precisión más clinómetro (Suunto 360PC/360R). Dado que estos dispositivos usados para medir el acimut son magnéticos, se deben corregir las lecturas por declinación magnética. En casos más complicados o donde se requiera mayor precisión, se emplearán equipos de estación total, junto a planos arqueológicos georreferenciados. Se estima que la muestra total esté conformada aproximadamente por 50 edificios. Como referencia se utilizarán recintos principales, fogones ceremoniales, escaleras principales, muros frontales, posteriores y laterales. Las medidas de acimut obtenidas en el sistema de coordenadas UTM WGS84 serán convertidas a medidas de acimut en el sistema de coordenadas geográfico empleando el ángulo de convergencia descrito por \cite{Thom1967}.

\subsection{Análisis de datos}

Los datos de acimut y altura del horizonte en la dirección medida nos permiten encontrar las coordenadas equivalentes en el sistema de coordenadas ecuatorial. En particular la coordenada declinación ($\delta$) permite establecer una relación directa con algún objeto del cielo. Sin embargo, obtener $\delta$ implica la necesidad adicional de corregir las medidas de \textit{h} por el efecto de la refracción atmosférica \citep{Aveni2001,Schaefer2000}. A esto se suma los efectos de la extinción atmosférica en el caso de estrellas que se encuentren muy bajas sobre el horizonte \citep{Schaefer1986}.
Los histogramas de $\delta$ sugieren muy claramente una posible intencionalidad (una alta correlación) en la orientación de las estructuras estudiadas hacia algún objeto del cielo. Además, estos diagramas resultan independientes de la ubicación geográfica y topografía local. Para construir estos histogramas de declinación es recomendable usar un “kernel” de suavizado apropiado para cada valor de $\delta$, lo que da origen a una "distribución de densidad kernel" (DDK). Luego, todos estos productos (DDKs) se suman para dar la DDK final de nuestros datos \citep{Weglarczyk2018}. Para poder afirmar que una medición es significativa, emplearemos una frecuencia relativa normalizada que fija la escala de nuestras DDKs o histogramas: se divide el número de ocurrencias de un valor dado por el valor promedio de ocurrencias de la muestra. Esto equivale a comparar con el resultado de una distribución uniforme en $\delta$ del mismo tamaño que la muestra de datos y con un valor igual al del promedio de los datos. Se espera que al graficar los datos así procesados surjan picos en ciertos valores de $\delta$ con amplitud por encima de cero, idealmente por encima del valor 3 (que tomamos como 3$\sigma$). 
En adición, se utilizará el test de confiabilidad Kolmogorov-Smirnov para analizar la dependencia de nuestro histograma frente a un conjunto de distribuciones aleatorias de declinación. Detalles sobre los diagramas de declinación y test de confiabilidad pueden encontrarse, por ejemplo, en \cite{Gonzalez-Garcia2021}, \cite{Paolo2020} y \cite{Gangui2016a}. A partir del análisis de nuestros datos podremos ver qué declinaciones se revelan más prominentes, si destacan una orientación principalmente solar o si seleccionan un asterismo o estrella particular, lo que dará paso al análisis e interpretación de los resultados teniendo en cuenta los aspectos culturales, rituales y sociales propios de la civilización Caral.


\bibliographystyle{baaa}
\small
\bibliography{bibliografia}
 
\end{document}